\documentclass[%
 aip,
 jmp,%
 amsmath,amssymb,
 reprint,%
]{revtex4-1}

\usepackage{inputenc}
\inputencoding{latin1}
\usepackage{textcomp}

\newcommand{\igraph}[2]{
\begin{center}\resizebox*{#1\textwidth}{!}{\includegraphics{#2}}\end{center}
}
\newcommand{\units}[1]{\relax\ifmmode{\ \mathrm{#1}}\else{\ \textrm{#1}}\fi}
\def\us{\units}
\newcommand{\degrees}{\textrm{\textdegree}}

\usepackage{subfigure}
\usepackage{enumerate}
\usepackage{graphicx}
\usepackage{dcolumn}
\usepackage{bm}

\begin{document}

Copyright (2014) American Institute of Physics. This article may be downloaded
for personal use only. Any other use requires prior permission of the author
and the American Institute of Physics.

The following article appeared in J. Appl. Phys. 115 203102 (2014) and may
be found at \url{http://dx.doi.org/10.1063/1.4878939}

\preprint{AIP/123-QED}

\title[Quasi-forbidden XRD composition determination]{Composition determination of quaternary GaAsPN layers from single XRD measurement of quasi-forbidden (002) reflection}

\author{J.-M.~Tilli}
 \email{juha-matti.tilli@iki.fi.}
\author{H.~Jussila}%
\affiliation{Department of Micro and Nanosciences, Aalto University, P.O. Box 13500, FI-00076 Aalto, Finland}

\author{K.~M.~Yu}
\affiliation{Materials Sciences Division, Lawrence Berkeley National Laboratory, Berkeley, CA 94720, USA}
\author{T.~Huhtio}%
\author{M.~Sopanen}%
\affiliation{Department of Micro and Nanosciences, Aalto University, P.O. Box 13500, FI-00076 Aalto, Finland}

\date{\today}

\begin{abstract}
GaAsPN layers with a thickness of 30nm were grown on GaP substrates with
metalorganic vapor phase epitaxy to study the feasibility of a single X-ray
diffraction (XRD) measurement for full composition determination of quaternary
layer material. The method is based on the peak intensity of a quasi-forbidden
(002) reflection which is shown to vary with changing arsenic content for
GaAsPN. The method works for thin films with a wide range of arsenic contents
and shows a clear variation in the reflection intensity as a function of
changing layer composition. The obtained thicknesses and compositions of the
grown layers are compared with accurate reference values obtained by Rutherford
backscattering spectroscopy combined with nuclear reaction analysis
measurements. Based on the comparison, the error in the XRD defined material
composition becomes larger with increasing nitrogen content and layer
thickness. This suggests that the dominating error source is the deteriorated
crystal quality due to the nonsubstitutional incorporation of nitrogen into the
crystal lattice and strain relaxation. The results reveal that the method
overestimates the arsenic and nitrogen content within error margins of about
0.12 and about 0.025, respectively.  
\end{abstract}

\maketitle

\section{Introduction}
GaAs$_{y}$P$_{1-x-y}$N$_{x}$ is an interesting material with many potential
applications. The incorporation of As and N into this material reduces the band
gap and changes the indirect band gap of GaP to a direct one, enabling the fabrication
of optoelectronic components on GaP substrates. As the lattice constant of GaP is
very close to that of silicon, with a high-quality buffer layer it is also
possible to fabricate these components on top of silicon platforms. As a matter of fact, an electrically pumped semiconductor laser of
GaAsPN quantum wells monolithically integrated on silicon has already been
demonstrated\cite{SiLaserAPL}. Furthermore, GaAsPN has recently drawn
increasing amount of attention also in the solar cell research
community. For instance, GaAsPN has been proposed as a
material to be used in the fabrication of a silicon tandem solar
cell\cite{almosni2013evaluation}. In addition to this, the nitrogen present in this material
splits the conduction band into two\cite{yu2006multiband}, potentially allowing the fabrication of an
intermediate-band solar cell\cite{IBSC}.

However, the composition determination of quaternary materials remains a
challenge. X-ray diffraction (XRD)\cite{XRDBartels,XRDLDS} is a commonly used
method to determine the composition of compound semiconductors. The method is
based on determining the location of the XRD peak, but it does not directly
allow full composition determination of quaternary semiconductors. As an
example of the challenges, the composition of the GaAsPN layer in the aforementioned silicon laser
was ''estimated by the strain state in line with various test structures
investigating the incorporation behavior of the different elements in
MOVPE''\cite{SiLaserAPL}. A second free variable, such as determination of the
band gap energy, is typically required for full composition determination. 

Previously, it has been demonstrated that the peak intensity of a
quasi-forbidden reflection varies strongly as a function of material
composition for InAlAsSb\cite{Quaternary}.  In this work, it is shown that the
peak intensity of the quasi-forbidden (002) XRD reflection of GaAsPN also varies
strongly with arsenic content. This factor is held as a second free variable
and used to determine the material composition. The obtained compositions and
layer thicknesses are compared to accurate reference values obtained with Rutherford
backscattering spectroscopy (RBS) combined with nuclear reaction analysis (NRA) measurements.

\section{Experimental details}

The samples studied in this paper were grown with metalorganic vapor phase
epitaxy (MOVPE) at atmospheric pressure. The used precursors were
trimethylgallium (TMGa), tertiarybutylarsine (TBAs), tertiarybutylphosphine
(TBP) and dimethylhydrazine (DMHy) for gallium, arsenic, phosphorus and
nitrogen, respectively. The GaAsPN layers were grown directly on top of GaP
(001) substrates. The growth temperature was $600\us{\degrees{}C}$ and the
samples were subsequently annealed for 5 minutes at $750\us{\degrees{}C}$.
Nominally 130 nm thick GaAsPN layers (samples \#6 - \#9) were grown with
similar flows as the 30 nm thick layers (samples \#5, \#4, \#1 and \#2,
respectively) for RBS/NRA measurements. The used V/III ratio for the samples
was about 100 but varied slightly for each sample. The information about the
sample growth relating to the used V/III, TBAs/V and DMHy/V ratios is found in
Table \ref{combinedtable}.

The strain state, crystallographic plane tilt, composition and thickness of the
grown layers were studied with a commercial Philips X'Pert Pro MRD
diffractometer. The measurement was performed in a high resolution mode using a
Ge (220) monochromator and an X-ray mirror at the incident beam side and an
analyzer crystal at the diffracted beam side. We also performed similar
measurements with an open detector at the diffracted beam side, but the
measured curve did not correspond to the simulated curve that well and the
accuracy of the method was worse.

Results from the XRD measurements were obtained with fitting analyses made with
a custom XRD curve fitting software\cite{tilli2007puolijohderakenteiden}
that utilized dynamical diffraction theory. The calculated electric
susceptibilities and XRD curves and values obtained with the fitting analysis
with the custom software have been compared with other software and found to be
in good agreement. It should be noted that the used deviation parameter
formula\cite{BartelsHornstraLobeek} may not be accurate with highly mismatched
epitaxial structures, so we tested simulated curves of a GaAsPN layer against
Sergey Stepanov's X-ray server at \url{http://x-server.gmca.aps.anl.gov/},
which uses a recursive matrix approach of dynamical X-ray
diffraction\cite{RMA}. The simulated curves were almost identical, and the
intensity of the thin film peak when compared to the substrate peak was the
same. Note that any XRD curve fitting software works with the used quaternary
composition determination method, as long as the value of the fitting error is
exposed to the user. Table \ref{params} lists the Poisson's ratios,
lattice constants and Debye-Waller $B$-factors of different elements required
for calculation of the simulated XRD curve. Note that the $B$-factors for P and
N were missing and a value of 0 was used instead. However, as the values for
the Debye-Waller factor, i.e.  $\exp(-Bs^2)$, for Ga and As are 0.984 and
0.981, respectively, the effect of the Debye-Waller $B$-factor on the
calculated electric susceptibility is small and therefore the missing values
for P and N can be assumed not to significantly alter the calculated electric
susceptibiliy value.

\begin{table}[tbp]
\caption{Simulation parameters used by the software.}
\begin{center}
\begin{tabular}{lll}
Material & parameter & value \\
\hline
GaP & Poisson's ratio & 0.3070 \\
GaAs & Poisson's ratio & 0.311 \\
cubic GaN & Poisson's ratio & 0.33 \\
GaP & lattice constant & 5.4505 Å \\
GaAs & lattice constant & 5.65368 Å \\
cubic GaN & lattice constant & 4.5034 Å \\
Ga & Debye-Waller $B$-factor & 0.46675 Å$^2$ \\
As & Debye-Waller $B$-factor & 0.55504 Å$^2$ \\
\end{tabular}
\end{center}
\label{params}
\end{table}

RBS and NRA measurements were performed for all the samples in order to get
accurate reference values for the layer thicknesses and compositions.
Compositions of the films were measured by channeling Rutherford backscattering
spectroscopy (c-RBS) together with nuclear reaction analysis (NRA). The
$^{14}$N($\alpha$,p)$^{17}$O reaction with a 3.72 MeV $^4$He$^{2+}$ beam was
used for detection of nitrogen. A 150 mm$^2$ passivated implanted planar
silicon (PIPS) detector with a 3$\times$12 mm slit was used to detect the
emitted protons at $135\degrees$ with respect to the incident beam. A
$25\us{\mu{}m}$ thick mylar foil was placed in front of the detector to absorb
the backscattered alpha particles. RBS spectra were also obtained
simultaneously at $165\degrees$ with another PIPS detector. Both RBS and NRA
measurements were carried out in random and $<$100$>$ axial channeling
directions. The fraction of substitutional nitrogen atoms in the films was
obtained by comparing the random and channeling yields of the RBS and the NRA
measurements.

\section{Theory}
As previously described, the use of the location of an XRD peak allows
composition determination of only ternary semiconductors, as only one free
composition-related variable, the lattice constant, can be deduced from the
peak location. Therefore, a second variable needs to be known to determine the composition of quaternary compounds. The second variable gathered by standard XRD diffractometer could for example be electron density determined by X-ray
reflectivity (XRR). However, it should be noted that it has been previously
estimated that the relative accuracy of the density determination for ALD-grown
Al$_2$O$_3$ on silicon is on the order of 3.5\%\cite{XrrAcc1,XrrAcc2}. Since Al$_2$O$_3$ on Si has a significantly better
electron density contrast when compared to GaAsPN on GaP, using electron density
determined by XRR as the second free variable with GaAsPN is expected to have
a significantly lower accuracy and hence does not seem to be feasible.

\begin{figure}[htbp]
\centering
\includegraphics[height=6cm]{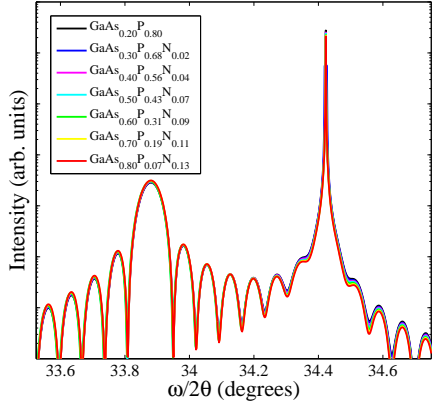}
\caption{Simulated XRD curves for the allowed (004) reflection of a 75 nm thick GaAs$_y$P$_{1-x-y}$N$_{x}$ layer with varying values of $y$ and with $x$ set so that
the lattice constant stays the same.}
\label{ascontent004}
\end{figure}

\begin{figure}[htbp]
\centering
\includegraphics[height=6cm]{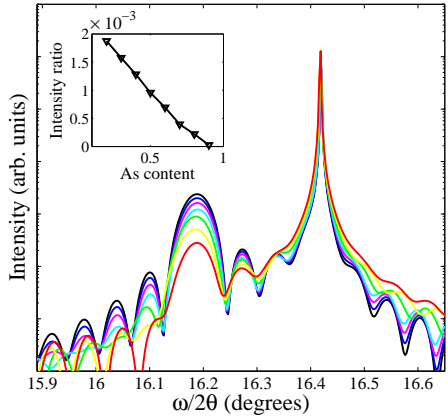}
\caption{Simulated XRD curves for the quasi-forbidden (002) reflection of a 75 nm thick GaAs$_y$P$_{1-x-y}$N$_{x}$ layer with varying values of $y$ and with $x$ set so that
the lattice constant stays the same.}
\label{ascontent002}
\end{figure}

Instead, the intensity of the XRD reflection is used in this work as the second
free variable for the definition of the material composition. To demonstrate
the effect, the XRD spectra of the allowed (004) and the quasi-forbidden (002)
reflection are simulated for a 75 nm thick GaAsPN layer on a GaP substrate as a
function of changing arsenic content. These XRD spectra are shown in figures
\ref{ascontent004} and \ref{ascontent002}. In the figures, the arsenic content
is varied with the lattice constant being kept constant by varying also the
nitrogen content.  Note that some of the curves are unphysical, \emph{i.e.}, it
is assumed that all of the layers are fully strained and that the nitrogen
content needed to keep the lattice constant the same is unrealistically high in
some of the curves. It can be seen from the figure that the intensity of the
allowed (004) reflection does not vary enough to be able to make deductions
about the composition based on the reflection intensity. However, for the
quasi-forbidden (002) reflection the intensity of the XRD peak varies strongly
as a function of the material composition. This occurs because for GaAs, the
scattering factors for Ga (31 electrons) and As (33 electrons) are very close
to each other. As a result of this, the interference from Ga and As atoms is
nearly but not completely destructive for the (002) reflection giving rise to a
non-zero structure factor. Additionally, replacing a certain small fraction of
As atoms (about 10\%) with P (15 electrons) causes the interference to be
completely destructive and replacing more of the As atoms with P causes the
interference to be less destructive. Note also that the same is true
for the (006) reflection.

The composition can be determined from the peak location and intensity using
two different methods. One possibility is comparing the integrated intensity of
the thin film peak with the substrate peak. This has been previously used to
determine the composition of quaternary InAlAsSb\cite{Quaternary}. Another
possibility is with an automatic fitting analysis. The advantage of the fitting
analysis over the integrated intensity method is that the fitting analysis works
for materials having a lattice constant closely matched to the substrate, as
the peaks of the substrate and the grown layer can be fused together. Such
was the case in our sample \#5. 

The available curve-fitting software typically allows fitting only one composition-related
parameter. Here a semiautomatic fitting analysis was used. The As content was set to a fixed trial value and the fitting analysis was performed with the
fitting parameters being the N content, the layer thickness, the
intensity normalization factor and a correction offset for the diffraction angle.
The minimum possible value of the fitting error was recorded for every trial As content and the value which minimizes the fitting error
was manually determined by drawing a graph of the minimum possible values of
the fitting error as a function of the As content. The used fitting error was obtained by calculating the 2-norm in logarithmic space.

It was found that using the peak intensity is not as accurate as using the peak position in
determining the composition of the material. In particular, the following
sources of error may be present:
\begin{enumerate}[(i)]
\item Poisson-distributed photon counting noise. This can result in the
      determined arsenic content being either too high or too low. This error
      source
      is random and can be reduced by using longer photon counting times.
      The magnitude of this error source may be estimated by performing
      the measurement multiple times and doing the fitting analysis for each
      measurement separately.
\item If the crystal planes of the thin film are tilted, the true peak
      may be at a different $\omega$ or $\psi$ angle relative to the peak of
      the substrate. This results in the peak observed in an 1-axis
      $\omega-2\theta$ scan being weaker than the true peak. The
      result is that the determined arsenic content is too high.
\item If the crystal quality of the layer is not good, the peak will be
      weaker than predicted by the theory. This error source will
      also result in the determined arsenic content being too high.
\item The layer may be partially relaxed. In this case, assuming
      the layer is compressively strained, the determined
      nitrogen level is too high, as both relaxation and an increase
      in nitrogen content tend to move the peak to a higher $\theta$
      angle. The relaxation may be studied with
      asymmetric reflections and the relaxation can be eliminated by
      ensuring that the layer thickness is below the critical thickness.
\item Inaccuracies in the used atomic scattering factors can also affect
      the results. As long as the used atomic scattering factors are
      reasonably close to the real values, the effect of the peak intensity
      varying as a function of the arsenic content can be seen. However, if
      they are slightly incorrect, the magnitude
      of the effect in simulations will change.

\item Inaccuracies in the Poisson's ratios and the lattice constants
      (Table \ref{params}) can affect the results, but they affect mainly
      the determined N content and not the electric susceptibilities which
      determine the peak intensity and thus the As content. Theoretically,
      inaccuracies in the Debye-Waller $B$-factors might also affect the
      calculated electric susceptibilities somewhat, but as $\exp(-Bs^2)$ is
      over 0.98 for the values of $s$ we used, the effect of the Debye-Waller
      $B$-factors is expected to be insignificant.

\end{enumerate}

It should also be noted that there is crosserror between the determined arsenic
content and the determined nitrogen content. If, \emph{e.g.}, due to one of the
aforementioned error sources the arsenic content determined from the peak
intensity is too high, this would move the peak to a lower $\theta$ angle
without an increase in the nitrogen content. The fitting analysis will then find
also the nitrogen content that is too high.

\section{Results and discussion}

\subsection{Full composition determination from a single XRD measurement}\label{fullxrd}
\begin{figure}[htbp]
\igraph{0.5}{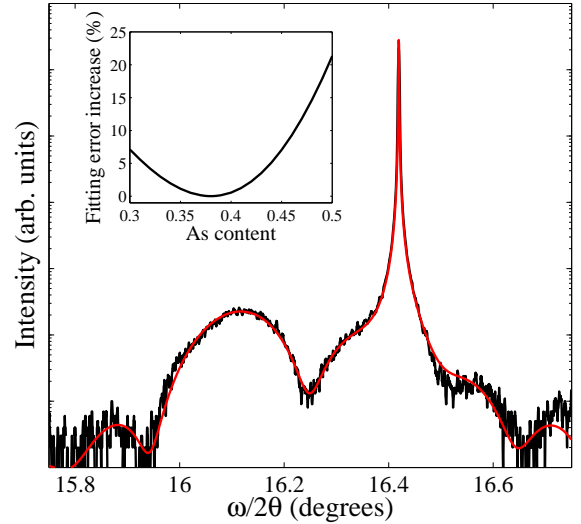}
\caption{First XRD measurement of sample \#1 and the fitted simulated curve. The inset shows how much the fitting error as a function of the As content is increased from the minimum fitting error value.}
\label{001_002_hq2_plot}
\end{figure}

The full compositions of the samples were determined with a single XRD scan of the (002)
reflection. A simulated model was fitted to the measured data with the
method presented earlier. Excellent fits were obtained which is illustrated in figure
\ref{001_002_hq2_plot} for sample \#1. The inset of figure \ref{001_002_hq2_plot} shows the obtained relative fitting error as a function of the trial
As content. The relative fitting error, which is presented with respect to the minimum fitting error, changes significantly and shows a clear minimum with the As content of 38 \%. Therefore, with the measurement of a single XRD curve of the quasi-forbidden (002) reflection, it is possible with high precision to unambigously define the layer composition of the quaternary material. 

\begin{figure}[htbp]
\igraph{0.5}{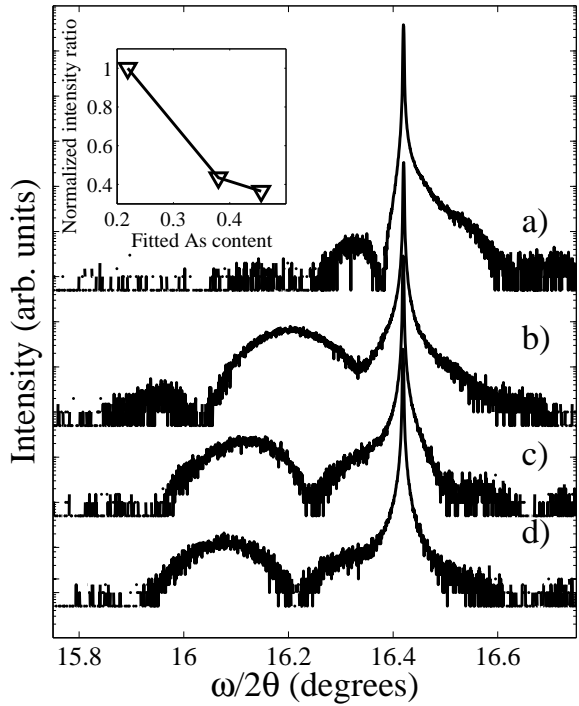}
\caption{XRD curves of samples (a) \#5, (b) \#4, (c) \#1, and (d) \#2. The inset shows the intensity ratio for the thin film peak and the substrate peak as a function of the fitted arsenic content, normalized so that the maximum obtained value is 1.}
\label{intensitycompare}
\end{figure}

The fits to the other measurements (not shown) were equally good and other
measurements had similarly a distinct variation of the fitting error as a function of the
arsenic content. The measured XRD curves from samples \#1, \#2, \#4 and \#5 are
shown together in figure \ref{intensitycompare} illustrating the variation
of the peak intensity for samples with different composition.
The XRD curve of sample \#3 having a similar arsenic content as sample \#2 is
not shown. It should be noted that as sample \#5 had no
arsenic, the peak of the layer is on the right side of the substrate and is
fused together with the substrate peak in this case. The small peak seen on the
left is an interference effect. The inset of figure \ref{intensitycompare}
shows the intensity ratios of the layer and the substrate peaks and clearly
demonstrates that the reflection intensity variation seen in simulations
(figure \ref{ascontent002}) occurs also in the XRD measurements of real
samples. 

The determined compositions and thicknesses of the grown layers are shown in
Table \ref{combinedtable}. It should be noted that the determined nitrogen
content seems to vary a lot depending on the arsenic content even though the
DMHy/V ratios used for all of the samples were similar. In addition, one
interesting observation is that no difference could be seen in the nitrogen
contents of samples \#2 and \#3 grown with different DMHy/V flows. This can be
explained by the fact that using the peak intensity is not as accurate in
determining composition as using the peak location and that there is the
previously mentioned crosserror between the determined arsenic and nitrogen
contents. This fact can be illustrated by forcing the arsenic content
of sample \#2 to be 0.439, the same as sample \#3. This causes the fitting
analysis to find a nitrogen content of 0.027 for sample \#2 in contrast
to the value of 0.031 which was obtained using the method described in this paper. In addition to this, it is
observed that the fitting analysis of the samples \#5 and \#6 grown with no
arsenic flow gives an arsenic content larger than zero which is unrealistic
because it is known that these grown layers cannot contain any arsenic.
Theoretically, some As could be present in the reactor from previously grown
samples, but our RBS measurements demonstrate that these layers did not contain
any arsenic. Additionally, it would be very unlikely that so much As would be incorporated into a grown sample from impurities present in the reactor, so the determined As content is due to a systematic error source in the composition determination method and not due to the samples having real As.

It can also be seen that for most samples, XRD underestimates the layer
thickness. This is likely caused by the fact that the measured (002) XRD curves
of especially the 130 nm thick layers did not have notable interference fringes and
thus the layer thickness was determined from the width of the XRD peak.
It is likely that the crystal quality of these samples had been deteriorated due to increased strain which caused the XRD peaks to broaden and lead the fitting to reveal a lower thickness than the actual thickness.

\begin{table*}[htbp]
\caption{Growth parameters, determined thicknesses and As and N contents of the samples. The sample \#1 was grown once but measurements were performed twice, which is denoted with \textsuperscript{(a}.}
\begin{center}
\begin{tabular}{|l|lll|ll|ll|}
\hline
Sample & \multicolumn{3}{c}{Growth parameter} & \multicolumn{2}{c}{Thickness} & \multicolumn{2}{c}{Composition} \\
     & V/III & DMHy/V & TBAs/V & nominal & measured & $y_\textrm{As}$ & $x_\textrm{N}$ \\\hline
\#1 & 101 & 0.32 & 0.045 & 30 nm & 29.5 nm & 0.380 & 0.022 \\
\#1 \textsuperscript{(a} & & & & & 29.1 nm & 0.374 & 0.020 \\
\#2 & 103 & 0.32 & 0.061 & 30 nm & 27.7 nm & 0.456 & 0.031 \\
\#3 & 106 & 0.39 & 0.054 & 30 nm & 28.2 nm & 0.439 & 0.030 \\
\#4 & 100 & 0.30 & 0.020 & 30 nm & 29.6 nm & 0.219 & 0.004 \\
\#5 & 98 & 0.31 & 0.000 & 30 nm & 28.3 nm & 0.082 & 0.025 \\
\#6 & 101 & 0.32 & 0.000 & 130 nm & 132.5 nm & 0.229 & 0.058 \\
\#7 & 103 & 0.32 & 0.020 & 130 nm & 99.7 nm & 0.693 & 0.119 \\
\#8 & 101 & 0.32 & 0.046 & 130 nm & 62.3 nm & 0.787 & 0.120 \\
\#9 & 103 & 0.32 & 0.061 & 130 nm & 64.7 nm & 0.776 & 0.110 \\\hline

\end{tabular}
\end{center}
\label{combinedtable}
\end{table*}

\subsection{RBS and NRA measurements}

\begin{table*}[htbp]
\caption{RBS and NRA measurement results. $f_\textrm{sub}$ refers to the
substitutional fraction of N in the crystal lattice. The right-most column presents the N content of the layer determined by the fitting analysis such a way that the arsenic composition of the GaAsPN layer is fixed to the value determined by RBS.}
\label{rbsnraresults}
\begin{center}
\begin{tabular}{|l|ccc|cc|c|c|}
\hline
Sample & \multicolumn{6}{c}{RBS/NRA} & \multicolumn{1}{c}{XRD} \\
 & Thickness & $y_\textrm{As,RBS}$ & $x_\textrm{N,NRA}$ & $\chi_\textrm{min,GaAsP}$ & $\chi_\textrm{min,N}$ & $f_\textrm{sub}$ & $x_\textrm{N,XRD}$\\\hline
\#1 & 31 nm           & 0.31 & ? &&&& 0.007 \\
\#2 & 31.5 nm         & 0.34 & ? &&&& 0.006 \\
\#3 & 31 nm           & 0.32 & ? &&&& 0.004 \\
\#4 & 31 nm           & 0.22 & ? &&&& 0.004 \\
\#5 & ?               & 0    & ? &&&& 0.006 \\
\#6 & $\approx$120 nm & 0    & $\approx$0.016 & 0.06 & 0.19 & 0.86 & 0.008 \\
\#7 & 124 nm          & 0.2  & $\approx$0.013 & 0.09 & 0.16 & 0.92 & 0.012 \\
\#8 & 125 nm          & 0.31 & $\approx$0.008 &&& & 0.017 \\
\#9 & 120 nm          & 0.32 & 0.01           &&& & 0.012 \\\hline
\end{tabular}
\end{center}
\end{table*}

In order to estimate the accuracy of the XRD measurement results, RBS and NRA
measurements were performed for all the samples. The results of the measurements
are shown in Table \ref{rbsnraresults}. Note that NRA measurements on the thin
samples ($\approx$30 nm) were not performed because the N signals were too low
to determine the N content in these cases. The thicknesses of the films were
estimated by assuming an average atomic composition of the film from the
measured As content and, therefore, are accurate to about 5\%. For the thicker
GaAsPN layers, NRA results were fitted by the SIMNRA software
\cite{mayer1999simnra} using a thin InN thin film as a standard for N
quantification.  The accuracy of the N measurements in these cases is
within 10\%. Note that for a perfect crystal the GaAsPN minimum yield
($\chi_\textrm{min,GaAsNP}$), ratio of the channeling to the random yields,
should be on the order of 0.04. However, for sample \#6 and \#7
$\chi_\textrm{min,GaAsNP}$ are significantly higher than that. This
suggests that extended defect density in these samples (particularly sample
\#7) is significantly higher giving rise to a higher dechanneling of the ion
beam.

To investigate how well the RBS and NRA determined As and N contents
agree with XRD measurements, we determined the N contents with XRD analysis by
forcing the As content to be the RBS-determined value instead of using the described
quaternary fitting method. The results are also shown in the right-most column of Table \ref{rbsnraresults}. It
can be seen that for the samples \#6 and \#7 the XRD determined N contents are
smaller than the values determined by NRA. This is consistent with earlier
experiments\cite{RBSNRA} that the determined N content is too low, as Ga(As)PN may
not exactly obey Vegard's law. However, for samples \#8 and \#9 the XRD determined
N content is higher than the NRA determined value. This is consistent with partial
relaxation of the layers.

\subsection{Error sources}
The XRD defined material compositions deviate significantly from the reference values obtained by the RBS/NRA measurements. To explain the differences, the magnitude of different error sources potentially present in the XRD analysis was estimated.  

\subsubsection{Photon counting noise}\label{photoncountingnoise}

To get an estimate how the photon-counting noise and other measurement-related
inaccuracies affect the determined composition, the XRD curve for sample \#1
was measured twice and the fitting analysis was performed separately for each
measurement. From the two measurements, it can be seen that the precision of
the method has absolute errors of about 0.006 and 0.002 for As and N contents,
respectively. Thus, the photon counting noise is not a significant error source
for the analysis. This is the case even though the intensity of the GaAsPN peak
(below 100 counts) is significantly smaller than the intensity of $10^4$
reported in the previous study for InAlAsSb\cite{Quaternary}.
Additionally, measurement repeatability was simulated by using the
simulated curve from GaAs$_{0.380}$P$_{0.598}$N$_{0.022}$ as the fitting target
after adding simulated Poisson-distributed noise to it. This was done five
times and the determined compositions are listed in Table
\ref{photonnoise}. Thus, even though the statistics was not significant, i.e., we repeated the measurement only twice for
one sample and did the measurement repeatability test five times, the results suggest that the photon counting noise error source is not significant in the analysis. Note that there may always be also other measurement reproducibility error sources such as sample misalignment. 

\begin{table}[htbp]
\caption{Fitting results of photon counting noise simulations performed to simulated XRD curve of having a GaAs$_{0.380}$P$_{0.598}$N$_{0.022}$ layer as a fitting target.}
\begin{center}
\begin{tabular}{|l|ll|}
\hline
Fitting run & $y_\mathrm{As}$ & $x_\mathrm{N}$ \\
\hline
1 & 0.375 & 0.021 \\
2 & 0.377 & 0.022 \\
3 & 0.375 & 0.021 \\
4 & 0.374 & 0.021 \\
5 & 0.380 & 0.022 \\\hline
\end{tabular}
\end{center}
\label{photonnoise}
\end{table}

\subsubsection{Crystallographic tilt and strain}
It has been assumed in the XRD fitting analysis that the crystal planes parallel to the surface of the thin film are not tilted differently from the substrate and that the thin film is fully strained. Either crystal plane tilt or relaxation can result in systematic errors in the measurement.
Relaxation occurs for all layers of material different from the substrate when
the critical thickness is exceeded and it has been shown that a GaP layer
grown on a misoriented silicon surface can relieve its strain energy by tilting
the crystal planes\cite{MisorientedTilt}. Therefore, it is important to check
that the layer is fully strained and that the crystal planes are not tilted,
although it should be noted that crystallographic tilt is extremely improbable
when the layer is grown on a nominally exactly cut wafers. The tilting of the crystal planes was studied with a reciprocal space map of
the symmetric (004) reflection. The reciprocal space maps (not shown) indicate
that for all of the 30 nm thick samples, the true thin film XRD peak occurs at
the same $\omega-\theta$ offset angle as the substrate peak with an accuracy of
$\Delta\omega < 0.01\degrees$. Therefore, the crystal planes are not tilted.

\begin{figure}[htbp]
\igraph{0.4}{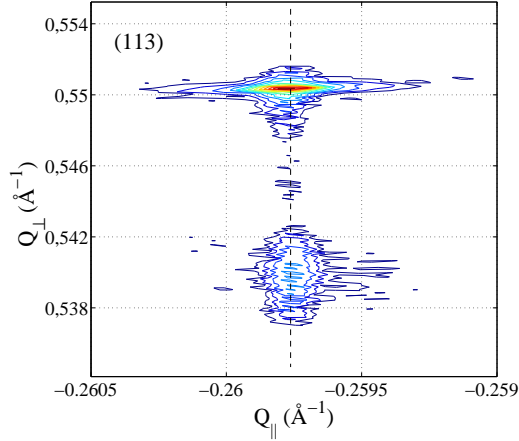}
\caption{Measured (113) reciprocal space map of the sample \#1. The horizontal
axis is the projection of the reciprocal space vector $Q$ to the axis parallel
to the sample surface and the vertical axis is its projection to the axis
perpendicular to the sample surface. Logarithmic scale has been used in the contour lines. The intensity range of the contour lines is from about 1 to about 8500 photons per second.}
\label{rsm113}
\end{figure}

The relaxation of the layer was studied with a reciprocal space map of the
asymmetric (113) reflection. All of the reciprocal space maps for the 30 nm
thick layers indicate that the layer peak occurs at the same value of the
diffraction vector component along the crystal plane, $Q_{||}$, as the
substrate peak and, thus, the layers are fully strained. The reciprocal space
map of the (113) reflection of sample \#1 is shown in figure \ref{rsm113}.
Reciprocal space maps were not measured for 130 nm thick samples.
However, assuming similar discrepancy between the experimentally
measured and theoretically calculated critical thickness as reported
for the GaP$_{0.98}$N$_{0.02}$ layers on GaP substrate\cite{jussila2013evaluation}, the critical thickness
of the GaAsPN layer in samples \#6, \#7, \#8, and \#9 can be estimated to
be on the order of 175~nm, 140~nm, 60~nm and 60~nm, respectively.
Thus it is likely that at least samples \#8 and
\#9 were partially relaxed.

\subsubsection{Impact of the layer thickness}
As explained in chapter \ref{fullxrd}, for all of the nominally 130 nm thick samples the determined arsenic content
was much higher than for the 30 nm thick samples grown with similar precursor
flows. This can be explained by the lower crystalline quality of the thicker
films due to increased lattice strain as the epilayer in none of the samples
was lattice matched to the GaP substrate. The presence of lattice strain
creates crystalline defects which causes the XRD peak to be weaker and broader
than that predicted by theory. This is in agreement with the high
channeling yield shown for these layers. This results in several effects:
\begin{enumerate}[(i)]
\item The determined arsenic content is too high, because high arsenic content
      and a low crystalline quality both result in a weak XRD peak.
\item The determined nitrogen content is too high due to two reasons: the
      layer may be relaxed, and there is a crosserror between the determined
      arsenic content and the determined nitrogen content. Additionally,
      relaxation may explain the broad and weak peak shapes observed and
      the determined low thickness values.
\item The determined thickness is too low, as the XRD curves did not have
      notable interference fringes and, thus, the layer thickness was determined
      from the width of the XRD peak. Relaxation causes the peaks
      to be broader than predicted by the dynamical diffraction theory,
      and thus the determined thickness will be too low.
\end{enumerate}

Based on this discussion, it can be concluded that the aforementioned error
sources are the smallest if the layer is as thin as possible.  However, at the
same time there is a trade-off because if the layer is very thin, the XRD peak
is wide and weak, which makes fitting analysis harder. The set of 130nm
thick samples had bad structural quality due to reasons explained before, and therefore, is not very useful from
the XRD measurement point of view.

\subsection{Failure of the simulation to reproduce the experimental intensity}

\begin{figure}[htbp]
\igraph{0.5}{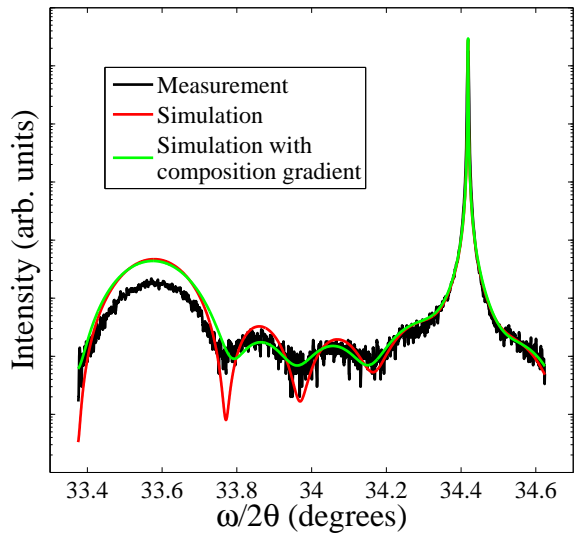}
\caption{XRD measurement of the (004) reflection of sample \#2 and the simulated XRD curve for GaAs$_{0.456}$P$_{0.513}$N$_{0.031}$ layer. The intensity normalization factor was fitted only based on
the substrate peak region. The curves have no vertical offset. It can be observed that the thin film peak is weaker than that predicted by dynamical diffraction theory.}
\label{002_004}
\end{figure}

From the comparison between the XRD and RBS/NRA results, it is known that the
XRD based method overestimates the arsenic content. Therefore, it seems to be
the case, simply judging from this fact, that the XRD peaks are weaker than the
theory predicts. To demonstrate this issue the (004) XRD reflection of sample
\#2 and the simulated XRD curve are shown in figure \ref{002_004}. Note that
the intensity of this reflection does not depend on the material composition.
It can be observed that the angular region in the measured XRD curve that
originates from the crystal planes of the GaAsPN epilayer produces a weaker
intensity than what the theory predicts. In the previous study of InAlAsSb
alloy\cite{Quaternary}, this observation has been called ''failure of the
simulation to reproduce the experimental intensity'' with no possible
explanations given other than inaccuracies in the used atomic scattering
factors. However, it can be deduced based on the comparison between the
different methods that the arsenic content does not seem to be off by the same
amount in all of the samples. Therefore, the weak intensity is not likely
originating from errors in the atomic scattering factors.

Another possible reason for the weakness of the peaks might be a
compositional gradient inside the GaAsPN layer. Note that all the samples were annealed in-situ for 5
minutes to enhance the substitutionality of N into the alloy. During the
anneal, however, no DMHy was flown inside the reactor which may cause that due
to desorption the N content of the layer can be lower near the surface of the
layer. Therefore, this hypothesis was tested fitting a sample model with two
layers of GaAsPN with one having a smaller N content than the other. A
simulation for the two layer model is also shown in figure
\ref{002_004}. It can be seen that in this case the intensity of the interference fringes agrees better to the experimentally measured XRD curve but the intensity level of the XRD peak of the layer is still similarly too high. Thus, a compositional gradient is
not the only source of discrepancy between the measurements and the dynamical
theory, and therefore, we believe that the most straightforward explanation for the weakness of the experimentally measured XRD peak is the presence of crystalline defects inside the GaAsPN layer.

A possible explanation for why the systematic error here is larger than
previously found is that nitrogen atoms can occupy interstitial sites in the
crystal lattice. For instance, it has been previously determined by RBS studies
combined with NRA measurements that with the nitrogen content of GaPN
increasing from 0.017 to 0.04 the substitutionality of the nitrogen decreases
from 0.91 to an unresolved value due to deteriorated crystal
quality\cite{RBSNRA}. Thus, it is likely that the samples studied in this work
contain nitrogen related point defects as well. In addition, it should be noted
that for a layer with a nitrogen content of 0.036, an incorrect value of 0.02
was obtained by XRD using Vegard's law. This difference was also explained by
different nitrogen configurations other than substitutional incorporation of
nitrogen into the crystal lattice affecting differently to the lattice
constant.\cite{RBSNRA} The samples used in this study were grown with the same
MOVPE apparatus.  Furthermore, it should be noted that it has also been
observed for other dilute nitride materials as well that the lattice constant
may deviate significantly from Vegard's law due to non-substitutional
incorporation of N into the crystal lattice\cite{NonSubstitutional}.

The interstitial N atoms, depending on their configuration, \emph{i.e.}, isolated
interstitials or split interstitials or other vacancy-interstitial complexes may
change the lattice constant in a different way and also additionally cause
strain fields around them, which affects the reflected intensity. In the
models used it is assumed that no nitrogen atoms occupy interstitial sites,
which is false in the real life. The substitutional fraction which was
determined for samples \#6 and \#7 gives further support to the hypothesis that
crystalline defects are the cause of the weakness of the XRD peaks. Based on
the determined substitutional fraction, it can be seen that part of the
nitrogen atoms are non-substitutionally incorporated into the crystal lattice.
More support to this explanation was given by growing a sample with a zero
arsenic content and a significantly higher nitrogen content. The determined
arsenic content using the new method increased even further for this sample even though the
sample contained no arsenic. In this case the agreement between the simulated
curve and the measured curve was significantly worse. Thus, the systematic error
increases with increasing nitrogen content and seems to be higher than the
error due to measurement repeatability.

For the 30nm thick samples, we can estimate the level of systematic
error from the comparison between the RBS and the XRD measurements. The level
of systematic error for the arsenic content is 0.12. The systematic error in
nitrogen content can be estimated by forcing the RBS-determined As content in
XRD fitting analysis and determining the N content. For example, for sample \#2
if we assume the As content of 0.34 in the fitting analysis, we obtain an N
content of 0.006, so the level of systematic error in N content is 0.025.

\subsection{Methods to minimize inaccuracies}
There are at least a few ways to improve the inaccuracies caused by crystalline
defects. First, the ratio of the integrated intensity of the thin film peak to
the substrate peak is calculated for both the (002) and (004) reflections. The ratio of these ratios can be compared with simulations of different
As contents, and the As content for which the ratio of ratios for the
measurement matches the simulation is taken to be the correct As content. However, such methods work better with the integrated peak intensity method presented earlier\cite{Quaternary} and cannot be used easily with the fitting analysis based method, which is the method that was demonstrated in this
paper. 

A more accurate dynamical theory of XRD which takes into account the effect of
crystalline defects in the thin film would be a more comprehensive way to
approach the problem. However, developing such a new theory of XRD is out of
the scope of this work. It should be noted that the effect of the weakness of
the XRD peak could be accounted for by multiplying the X-ray electric
susceptibility of the thin film by a certain factor (the same for (002) and
(004) reflections), which tends to make the thin film peak weaker. Such a
factor could be determined by fitting analysis from a measured (004) XRD
curve. However, such a multiplicative factor is not consistent with
the dynamical X-ray diffraction theory so the proposed solution is not
completely satisfactory. It can be speculated that by accounting this
''failure of the simulation to reproduce the experimental intensity'' error
source in the fitting procedure the accuracy of the method would approach the
limits explained in the chapter \ref{photoncountingnoise}. Such methods to
improve the inaccuracies presented here have not been studied here in detail
and require further work.

\section{Conclusions}

It was shown that the intensity of the quasi-forbidden (002) reflection of
GaAsPN varies strongly with arsenic content. Samples with thicknesses of 30 nm
and 130 nm and different compositions were grown on GaP substrates and were
measured with the designed XRD-based method and compared to accurate reference
values obtained by RBS combined with NRA measurements. It was observed that the
XRD-based method could unambigously determine material composition from a single scan of the
(002) reflection with high precision, but the accuracy of the method
for this set of samples was not good. Based on the comparison between the
different measurement methods, the error in the XRD defined material
composition became larger with increasing nitrogen content and layer thickness.
This suggested that the dominating error source was the deteriorated crystal
quality due to the nonsubstitutional incorporation of nitrogen into the crystal
lattice and strain relaxation affecting the reflection intensity unexpectedly.
More support for the conclusion that crystal quality was deteriorated
was obtained from the high channeling yield of the RBS/NRA measurements. We
also could find evidence of a compositional gradient of N from the fringe
modulation in a measurement of the (004) reflection, but this did not explain
the weakness of the main XRD peak. For 30 nm thick films the systematic error
limits were about 0.12 for the determined arsenic content and 0.025 for the
determined nitrogen content. It was proposed that the accuracy of the XRD-based
method can be significantly further enhanced possibly to the limit set by
typical error sources (photon counting noise, sample misalignment,
\emph{etc.}). This, however, requires a XRD fitting software designed in such a
way that it is capable to take into account deteriorated crystal quality. In
addition, a similar method could be used, in theory, for many other quaternary
compound semiconductor materials.

However, when applying the presented method for quaternary
composition determination, care should be taken to ensure that the layers have
a sufficient crystalline quality. Note that the simple method of determining ternary
layer composition from the location of the XRD peak works well for samples of
even low crystalline quality, but the same is not true for determining
quaternary material composition from the location and the intensity of the XRD
peak. The crystalline quality can be \emph{e.g.} checked by performing
measurements for the (004) reflection and ensuring that the dynamical theory
reproduces the experimental peak intensity.


%

\end{document}